\documentclass[twocolumn,pra,aps,showpacs]{revtex4}
\usepackage{amsfonts}
\usepackage{amssymb}
\usepackage{epsfig}
\usepackage{amsbsy}
\usepackage{amssymb}
\usepackage{amsmath}
\usepackage{graphicx}
\usepackage{graphics}

\begin{document}

\title{Multimode Phonon Cooling via Three Wave Parametric Interactions with
Optical Fields}
\author{G. S. Agarwal}
\affiliation{Department of Physics, Oklahoma State University, Stillwater, Oklahoma
74078, USA}
\author{Sudhanshu S. Jha}
\affiliation{UM-DAE Centre for Excellence in Basic Sciences, University of Mumbai
Vidyanagari Campus, Mumbai 400098, India}
\date{\today }

\begin{abstract}
We discuss the possible cooling of different phonon modes via three wave
mixing interactions of vibrational and optical modes. Since phonon modes
exhibit a variety of dispersion relations or frequency spectra with diverse
spatial structures, depending on the shape and size of the sample, we
formulate our theory in terms of relevant spatial mode functions for the
interacting fields in any given geometry. We discuss the possibility of
Dicke like collective effects in phonon cooling and present explicit results
for simultaneous cooling of two phonon modes via the anti-Stokes up
conversions. We show that the bimodal cooling should be observable
experimentally.
\end{abstract}

\maketitle


\section{Introduction}

The cooling of nano mechanical mirrors by the process of optical
up-conversion has become a standard technique \cite{ar0,bk0,bk01}. However
the same technique does not quite work for cooling of Brilliouin acoustic
modes \cite{ar5} in a bulk material medium. In this case, one has to satisfy
constraints arising from the phase matching conditions and at the same time
avoid the generation of Stokes radiation. In the simplest case it is known
that the Stokes process dominates \cite{bk2,ar3}. The latter becomes
critical because a thermal distribution of phonons would give rise to the
generation of both Stokes (down conversion) and anti-Stokes (up conversion)
fields. The dominant Stokes process would lead to heating. Bahl et. al. \cite%
{ar5} solved this problem by using a sophisticated resonator so that the
incident pump and generated anti-Stokes fields were resonant with two
different resonator modes, but the generated Stokes field was nonresonant.
In addition, they found a suitable Brilliouin mode in the spherical
resonator geometry, so that the relevant phase matching condition was
satisfied. More generally phonons are known to have many different types of
dispersion relations \cite{ar1,ar4} which in turn determine the phase
matching conditions and therefore one has to examine the cooling of phonon
modes under more general conditions. It is also very interesting to study
the cooling of two or even several different phonon modes simultaneously,
via anti-Stokes upconversion. This is very relevant in the context of
applications in quantum information science when phonons are used as
carriers of quantum information. In the light of this, in this paper we
develop a more general framework for possible cooling of general phonon
modes.

As mentioned above, our main aim here is to present a general theory of
laser cooling of low frequency phonon modes in matter, via the anti-Stokes
up- conversion process of the inelastic light scattering. Although, our
formulation will begin with an overview of the known case of laser cooling
of a single longitudinal acoustic phonon mode via the anti-Stokes Brillouin
scattering process, it will go on to develop a general theory of cooling of
low frequency phonon modes of arbitrary polarization, whether these are
acoustic or optical phonons, and whether these are modes in an extended bulk
matter or in a confined geometry. In fact, the formulation has been kept
general enough to be applicable not only to phonons but to any type of low
frequency collective excitation modes in matter which can give rise to
inelastic light scattering (Raman scattering) via the modulation of the
linear optical dielectric function. The cooling via the three -wave
anti-Stokes up -conversion is possible in principle for any such mode ,
provided that it has a sufficiently long life time compared to the resonant
cavity anti-Stokes optical mode, and the corresponding Raman tensor coupling
of the mode is not too weak. We will also consider the possibility of
cooling more than one phonon mode simultaneously via the three-wave
up-conversion process, which has not been considered earlier even for the
case of longitudinal acoustic phonons \cite{ar5,ar2}.

The organization of the paper is as follows-In Sec II we introduce the basic
features of the phonon photon interaction, relevant to anti-Stokes
generation. We derive the basic equations for the coupled phonon and
electromagnetic fields. In Sec III we derive the equations for phonon and
photon fields in terms of the relevant resonator modes. The nature of the
phonon dispersion or its frequency spectrum, and its relevance for
phonon-photon interaction are discussed in Sec IV. In Sec V we describe
quantum Langevin equations and show how the single mode cooling emerges \cite%
{ar2}. In Sec VI we discuss collective effects in simultaneous cooling of
two modes. The numerical results for the simultaneous cooling of two modes
are given in Sec VII, with a discussion of these results.

\section{Basic Equations for Interaction of Electromagnetic Fields with
Vibrational Fields}

In this section, we present, for completeness, a first principle derivation
of the basic equations for the Stokes and anti-Stokes scattering from
vibrational modes. This basic description would enable us to formulate the
problem of more general three wave interactions in a resonator. For
simplicity, let us first consider the case of longitudinal acoustic waves in
matter, which give rise to the Brillouin scattering. The long wavelength
longitudinal acoustic waves can be described either in terms of density
fluctuations $\Delta \rho (\mathbf{r},t)$ or the displacement field $\mathbf{%
u}(\mathbf{r},t)$, which are related by
\begin{equation}
\frac{\Delta \rho (\mathbf{r},t)}{\rho _{0}}=-\nabla \cdot \mathbf{u}(%
\mathbf{r},t),  \label{e1}
\end{equation}%
where $\Delta \rho $, $\rho _{0}$ are, say, mass densities, and $\rho _{0}$
is the mean mass density of the medium. In terms of the displacement field $%
\mathbf{u}(\mathbf{r},t)$, the unperturbed Hamiltonian of the acoustic field
is given by
\begin{eqnarray}
H_{S} &=&\int d^{3}\mathbf{r}\mathcal{H}_{S},  \label{e2} \\
\mathcal{H}_{S} &=&\frac{1}{2}\rho _{0}\dot{u}^{2}+\frac{1}{2}\rho
_{0}v_{S}^{2}(\nabla \cdot \mathbf{u})^{2},  \label{e3}
\end{eqnarray}%
where $v_{S}$ is the sound velocity. Also, for longitudinal sound wave, $%
\rho _{0}v_{S}^{2}=B$, which is the inverse of the compressibility $c_{S}$.
The equation (\ref{e2}) leads to the wave equation for sound waves
\begin{equation}
\frac{\partial ^{2}\mathbf{u}}{\partial t^{2}}-v_{S}^{2}\nabla ^{2}\mathbf{u}%
=\frac{\mathbf{f_{\mathrm{ext}}}}{\rho _{0}},  \label{e4}
\end{equation}%
where $\mathbf{f_{\mathrm{ext}}}$ is external force if any. Because of the
sound wave, the electromagnetic polarization of the medium is modulated by
the density wave. The change $\Delta \mathbf{P}$ in polarization can be
written in terms of the dielectric function as
\begin{eqnarray}
\Delta \mathbf{P} &=&\Delta \left( \frac{\epsilon -1}{4\pi }\right) \mathbf{E%
}  \notag \\
&=&\frac{1}{4\pi }\left( \frac{\partial \epsilon }{\partial \rho }\right)
_{0}\Delta \rho \mathbf{E}=\frac{\rho _{0}}{4\pi }\left( \frac{\partial
\epsilon }{\partial \rho }\right) _{0}\frac{\Delta \rho }{\rho _{0}}\mathbf{E%
}  \notag \\
&=&-\frac{\gamma _{e}}{4\pi }(\nabla \cdot \mathbf{u})\mathbf{E,}  \label{e5}
\end{eqnarray}%
where
\begin{equation}
\gamma _{e}\equiv \rho _{0}\left( \frac{\partial \epsilon }{\partial \rho }%
\right) _{0}  \label{e6}
\end{equation}%
is the so called electrostrictive constant. For simplicity we consider an
isotropic medium. The interaction density (interaction between the sound
wave and optical fields) is thus given by \cite{bk1}:
\begin{equation}
\mathcal{H}_{\mathrm{int}}=-\frac{1}{2}\Delta \mathbf{P}\cdot \mathbf{E}=%
\frac{\gamma _{e}}{8\pi }(\nabla \cdot \mathbf{u})E^{2},  \label{e7}
\end{equation}%
so that the interaction Hamiltonian $H_{\mathrm{int}}=\int \mathcal{H}_{%
\mathrm{int}}(\mathbf{r})d^{3}\mathbf{r}$. In view of the volume
integration, Eq. (\ref{e7}) also be written as
\begin{equation}
\mathcal{H}_{\mathrm{int}}=-\frac{\gamma _{e}}{8\pi }(\mathbf{u}\cdot \nabla
)E^{2}.  \label{e8}
\end{equation}

From Eq.(\ref{e8}) we can obtain the force $f_{\alpha }$ acting on the
vibrational mode which would be $-\partial \mathcal{H}/\partial u_{\alpha }$%
, i.e.
\begin{equation}
\mathbf{f}=\frac{\gamma _{e}}{8\pi }\nabla E^{2}.  \label{e9}
\end{equation}%
The quantity $-\frac{\gamma _{e}}{8\pi }E^{2}$ can be identified as the
pressure due to electromagnetic waves. Using (\ref{e9}) the wave equation
for $\mathbf{u}$ become
\begin{equation}
\frac{\partial ^{2}\mathbf{u}}{\partial t^{2}}-v_{S}^{2}\nabla ^{2}\mathbf{u}%
-\Gamma ^{\prime }\nabla ^{2}\frac{\partial \mathbf{u}}{\partial t}=\frac{%
\gamma _{e}}{8\pi \rho _{0}}\nabla E^{2},  \label{e10}
\end{equation}%
where we have also introduced the damping parameter $\Gamma ^{\prime }$ for
sound waves.

The electromagnetic field $\mathbf{E}$ satisfies wave equation
\begin{equation}
\nabla ^{2}\mathbf{E}-\frac{\epsilon }{c^{2}}\frac{\partial ^{2}\mathbf{E}}{%
\partial t^{2}}=\frac{4\pi }{c^{2}}\frac{\partial ^{2}(\Delta \mathbf{P})}{%
\partial t^{2}},  \label{e11}
\end{equation}%
where $\epsilon $ is the linear dielectric function of the medium. In
writing the above equation, it has to be noted that we have ignored the
linear dispersion of the dielectric function (which leads to the
corresponding group velocity instead of the phase velocity) and the linear
absorption of optical fields. Later we will however add the linear
absorption term as a phenomenological parameter.

Using Eq. (\ref{e5}), Eq. (\ref{e11}) reduces to
\begin{equation}
\nabla ^{2}\mathbf{E}-\frac{\epsilon }{c^{2}}\frac{\partial ^{2}\mathbf{E}}{%
\partial t^{2}}=-\frac{\gamma _{e}}{c^{2}}\frac{\partial ^{2}}{\partial t^{2}%
}\left[ \left( \nabla \cdot \mathbf{u}\right) \mathbf{E}\right] ,
\label{e12}
\end{equation}

The analysis given above can be generalized to any type of vibrational mode
in an arbitrary medium, whether it is an acoustic or optical phonon mode, a
longitudinal or transverse mode in a bulk material, or whether it is a mode
in a confined geometry. We can expand $\epsilon $ as a Taylor series in the
relevant phonon displacement field $\mathbf{Q}(\mathbf{r},t)$ whence
\begin{equation}
(\Delta \mathbf{P})_{i}=\sum_{jk}\frac{1}{4\pi }\frac{\partial \epsilon _{ij}%
}{\partial Q_{k}}E_{j}Q_{k}=\sum_{jk}R_{ijk}E_{j}Q_{k}.  \label{e13}
\end{equation}%
where $R_{ijk}$ is the generalized Raman tensor for the mode $\mathbf{Q}$.
If more than one phonon mode is involved, then Eq. (\ref{e13}) is to be
summed over contributions from all the phonon modes. The interaction
Hamiltonian density now becomes
\begin{equation}
\mathcal{H}_{\mathrm{int}}=-\frac{1}{2}\sum_{ijk}R_{ijk}E_{j}E_{i}Q_{k}.
\label{e14a}
\end{equation}%
It should be emphasized here that we have introduced a new variable $\mathbf{%
Q}$ for the phonon displacement field for a general phonon mode in matter,
instead of the variable $\mathbf{u}$ used for the low frequency longitudinal
acoustic displacement field, on purpose. Both have the same dimensions of
length. But, whereas $\mathbf{u}$ is related to the density fluctuation via
Eq. (\ref{e1}), and $\nabla \cdot \mathbf{u}=0$ for transverse modes, $%
\mathbf{Q}$ represents the displacement field for any phonon mode, whether
it is transverse or longitudinal, or whether it is an acoustic mode or an
optical mode. In fact, in the plane-wave representation for the case of bulk
matter, for a mode with wave vector $\mathbf{q}$ and a given general
dispersion relation $\Omega \left( \mathbf{q}\right) $, the displacement
field $\mathbf{Q}$ satisfies the equation, $\frac{\partial ^{2}\mathbf{Q}}{%
\partial t^{2}}+\Omega ^{2}\left( \mathbf{q}\right) \mathbf{Q}=0$, with its
velocity $\mathbf{v}=\partial \Omega /\partial \mathbf{q}$. For the low
frequency longitudinal acoustic mode, $\mathbf{Q}$ will be the same as $%
\mathbf{u}$, with $\Omega \left( \mathbf{q}\right) =v_{S}q$. Also, it should
be noted that the dimensions of the generalized Raman tensor $R_{ijk}$
differ from the dimensions of $\gamma _{e}$ of Eq. (\ref{e6}) by a factor of
($1/$length).

In terms of unit polarization vectors, $\hat{e}^{\left( 1\right) }$, $\hat{e}%
^{\left( 2\right) }$ and $\hat{e}^{\left( Q\right) }$ for the three
interacting modes, namely, the two optical modes and one phonon mode,
respectively, relevant to the process being considered, one can also define
a scalar effective Raman coupling constant for a bulk material, separately
for the Stokes and the anti-Stokes interaction. For example, if $2$ labels
the anti-Stokes optical mode, for the anti-Stokes process one has the
effective bulk coupling constant%
\begin{equation}
R^{\left( Q\right) }=\sum_{ijk}R_{ijk}\hat{e}_{i}^{\left( 2\right) \ast }%
\hat{e}_{j}^{\left( 1\right) }\hat{e}_{k}^{\left( Q\right) }.  \label{e14b}
\end{equation}%
For the special case of the long wavelength bulk longitudinal acoustic mode
of wave vector $\mathbf{q}$, frequency $\omega _{m}$ and velocity $v_{S}$,
described by Eqs. (\ref{e10}) and (\ref{e11}), one has%
\begin{equation}
4\pi R_{ijk}=\text{i}\gamma _{e}\delta _{ij}q_{k};%
\begin{array}{cc}
&
\end{array}%
4\pi R^{\left( Q\right) }=\text{i}\gamma _{e}q=\text{i}\gamma _{e}\frac{%
\omega _{m}}{v_{S}},  \label{e14c}
\end{equation}%
which follows on comparing Eqs. (\ref{e14a}) and (\ref{e14b}) with Eq. (\ref%
{e8}).

\section{Interaction between Electromagnetic Fields and phonons in a
Resonator}

The coupled equations (\ref{e10}) and (\ref{e12}) are general enough to
describe a variety of situations. Since we are interested here in cooling
issues, we consider the case involving modes in resonators. We also
specialize to the case of the anti-Stokes scattering. Let the
electromagnetic field $\mathbf{E}$ consist of waves $\mathbf{E}_{1}$ and $%
\mathbf{E}_{2}$ at the frequency $\omega _{1}$ (called pump), and $\omega
_{2}$ (anti-Stokes field). The frequencies $\omega _{1}$ and $\omega _{2}$
would be in the neighborhood of the resonator frequencies $\omega _{\mathrm{%
c1}}$ and $\omega _{\mathrm{c2}}$. Each resonator frequency would have
certain line width. It is assumed that $\omega _{i}-\omega _{\mathrm{c}i}$
is within the line width of the mode. We can then write the field as
\begin{equation}
\mathbf{E}_{i}(\mathbf{r},t)=\vec{\phi}_{i}\mathcal{E}_{i}(t)\mathrm{e}^{-%
\mathrm{i}\omega _{i}t}+c.c.%
\begin{array}{cc}
. &
\end{array}
\label{e15}
\end{equation}%
In our formulation, we will use the familiar slowly varying amplitude
approximation (SVAA) for finding the temporal variation of the field
amplitudes $\mathcal{E}_{i}(t)$. Here $\vec{\phi}_{i}$ (with $\nabla \cdot
\vec{\phi}_{i}=0$) is the spatial mode function for the electromagnetic mode
in the resonator at the frequency $\omega _{\mathrm{c}i}$ such that $\nabla
^{2}\vec{\phi}_{i}+\frac{\omega _{\mathrm{c}i}^{2}}{c^{2}}\epsilon _{i}\vec{%
\phi}_{i}=0$. The amplitude $\mathcal{E}_{i}(t)$ is a slowly varying
function of $t$. We use
\begin{eqnarray*}
\nabla ^{2}\mathbf{E}_{i}-\frac{\epsilon _{i}}{c^{2}}\frac{\partial ^{2}%
\mathbf{E}_{i}}{\partial t^{2}} &\cong &(\nabla ^{2}\vec{\phi}_{i}+\frac{%
\epsilon _{i}}{c^{2}}\omega _{i}^{2}\vec{\phi}_{i})\mathcal{E}_{i}\mathrm{e}%
^{-\mathrm{i}\omega _{i}t}+ \\
&+&2\mathrm{i}\frac{\omega _{i}\epsilon _{i}}{c^{2}}\frac{\partial \mathcal{E%
}_{i}}{\partial t}\mathrm{e}^{-\mathrm{i}\omega _{i}t}\vec{\phi}_{i}+c.c.
\end{eqnarray*}%
which on using $\omega _{i}^{2}-\omega _{\mathrm{c}i}^{2}\approx 2\omega _{%
\mathrm{c}i}(\omega _{i}-\omega _{\mathrm{c}i})=-2\omega _{\mathrm{c}%
i}\Delta _{i}$ reduces to
\begin{eqnarray}
&&\nabla ^{2}\mathbf{E}_{i}-\frac{\epsilon _{i}}{c^{2}}\frac{\partial ^{2}%
\mathbf{E}_{i}}{\partial t^{2}}  \notag \\
&\approx &2\vec{\phi}_{i}\frac{\epsilon _{i}}{c^{2}}\omega _{\mathrm{c}i}%
\left[ \mathrm{i}\frac{\partial \mathcal{E}_{i}}{\partial t}-\Delta _{i}%
\mathcal{E}_{i}\right] \mathrm{e}^{-\mathrm{i}\omega _{i}t}+c.c.,
\label{e16}
\end{eqnarray}%
$\Delta _{i}=\omega _{\mathrm{c}i}-\omega _{i}$. Note that we have kept only
the terms which are up to the first order in the time derivative or in the
frequency difference $\Delta _{i}$. We next write the phonon mode as
\begin{equation}
\mathbf{u}(\mathbf{r},t)=\vec{\psi}(\mathbf{r})u(t)\mathrm{e}^{-\mathrm{i}%
\omega _{m}t}+c.c.,  \label{e17}
\end{equation}%
where, the mode function $\vec{\psi}(\mathbf{r})$ has the dimensions of the
displacement field itself and $u(t)$ is the dimensionless amplitude of the
mode. Note that phonons are interacting with a thermal bath, and depending
on the strength of this interaction, the frequency spectrum of the
amplitudes $u(t)$ for the given mode of frequency $\omega _{m}$ will
represent the line width around the frequency $\omega _{m}$. When such
phonons interact with the external optical fields via the three-wave
interaction, this spectrum would be modified further. For longitudinal
phonons curl $\vec{\psi}(\mathbf{r})=0$. Using the slowly varying amplitude
approximation for the phonon mode amplitude also, the procedure that led to
Eq. (\ref{e16}) now leads to%
\begin{equation}
-\nabla ^{2}\mathbf{u}+\frac{\partial ^{2}\mathbf{u}}{v_{S}^{2}\partial t^{2}%
}\approx 2\frac{\vec{\psi}}{v_{S}^{2}}\omega _{m}\left[ -\mathrm{i}\frac{%
\partial u}{\partial t}\right] \mathrm{e}^{-\mathrm{i}\omega _{m}t}+c.c.%
\begin{array}{cc}
. &
\end{array}
\label{e18}
\end{equation}%
Next we need to work out the interaction terms. Since we are considering the
anti-Stokes process, $\omega _{1}+\omega _{m}\approx \omega _{2}$. This
relation has to be satisfied within the line widths of all three modes $%
\omega _{\mathrm{c1}}$, $\omega _{\mathrm{c2}}$ and $\omega _{m}$ and hence
all subsequent statements are to be understood within linewidths. Clearly a
term like $\mathcal{E}_{1}^{\ast }\mathcal{E}_{2}$ on the right hand side of
Eq.(\ref{e10}) will lead to the mode $\mathbf{u}$. Then the righthand side
of Eq. (\ref{e10}) is to be approximated by
\begin{equation}
\nabla E^{2}=\mathcal{E}_{1}^{\ast }\mathcal{E}_{2}\nabla \left( \vec{\phi}%
_{1}^{\ast }\cdot \vec{\phi}_{2}\right) \mathrm{e}^{\mathrm{i}\left( \omega
_{1}-\omega _{2}\right) t}+c.c.%
\begin{array}{cc}
. &
\end{array}
\label{e19}
\end{equation}%
Similarly for deriving equation, say, for $\mathcal{E}_{2}$ from Eq. (\ref%
{e12}), the right hand side is to be approximated by%
\begin{eqnarray}
&&-\frac{\gamma _{e}}{c^{2}}\frac{\partial ^{2}}{\partial t^{2}}\left[
\left( \nabla \cdot \mathbf{u}\right) \mathbf{E}\right]   \notag \\
&\cong &\frac{\gamma _{e}}{c^{2}}\omega _{2}^{2}\left( \mathcal{E}%
_{1}u\left( \nabla \cdot \vec{\psi}\right) \vec{\phi}_{1}\right) \mathrm{e}%
^{-\mathrm{i}\left( \omega _{1}+\omega _{m}\right) t}+c.c.%
\begin{array}{cc}
. &
\end{array}
\label{e20}
\end{eqnarray}%
On combining Eqs. (\ref{e16})-(\ref{e20}) and carrying out all the
simplifications, we get final equations for $\mathcal{E}_{1}$, $\mathcal{E}%
_{2}$ and $u$:%
\begin{eqnarray}
\frac{\partial \mathcal{E}_{2}}{\partial t} &=&-\kappa _{2}\mathcal{E}_{2}-%
\text{i}\Delta _{2}\mathcal{E}_{2}-\text{i}\frac{\chi _{2}}{\epsilon _{2}}%
\omega _{\mathrm{c}2}u\mathcal{E}_{1}\mathrm{e}^{\mathrm{i}\delta t},
\label{e21} \\
\frac{\partial \mathcal{E}_{1}}{\partial t} &=&-\kappa _{1}\mathcal{E}_{1}-%
\text{i}\Delta _{1}\mathcal{E}_{1}-\text{i}\frac{\chi _{2}^{\ast }}{\epsilon
_{1}}\omega _{\mathrm{c}1}u^{\ast }\mathcal{E}_{2}\mathrm{e}^{-\mathrm{i}%
\delta t},  \label{e22} \\
\frac{\partial u}{\partial t} &=&-\Gamma u-\text{i}\frac{\chi _{2}^{\ast }%
\mathcal{E}_{1}^{\ast }\mathcal{E}_{2}\mathrm{e}^{-\mathrm{i}\delta t}}{4\pi
\rho _{0}\omega _{m}\int \vec{\psi}^{\ast }\left( \mathbf{r}\right) \cdot
\vec{\psi}\left( \mathbf{r}\right) d^{3}\mathbf{r}},  \label{e23} \\
\delta  &=&\omega _{2}-\omega _{1}-\omega _{m},  \label{e24} \\
\chi _{2} &=&\frac{\gamma _{e}}{2}\int \left( \vec{\phi}_{2}^{\ast }\left(
\mathbf{r}\right) \cdot \vec{\phi}_{1}\left( \mathbf{r}\right) \right)
\left( \nabla \cdot \vec{\psi}\left( \mathbf{r}\right) \right) d^{3}\mathbf{%
r.}  \label{e25}
\end{eqnarray}%
Here $\chi _{2}$ is the effective three mode coupling which depends on the
overlap of all the modes. The integral in Eq. (\ref{e25}) is the traditional
phase matching integral if plane waves are used for all the three modes in a
bulk material. In writing Eqs. (\ref{e21})-(\ref{e25}), we have added
linewidth terms for all the modes. Note that $\gamma _{e}\left( \nabla \cdot
\mathbf{u}\right) $ is dimensionless and $\chi _{2}$ is dimensionless, the
term $\mathcal{E}^{2}$ has dimensions of energy and $\rho
_{0}v_{S}^{2}\left( \nabla \cdot \vec{\psi}\right) ^{2}u^{2}$ has dimensions
of energy density, keeping in view that $u$ and $\nabla \cdot \vec{\psi}$
are dimensionless. The term $\rho _{0}$ in Eq.(\ref{e23}) can be eliminated
by using the kinetic energy term $\frac{1}{2}\int \rho _{0}^{2}\left\vert
\mathbf{\dot{u}}\right\vert ^{2}d^{3}\mathbf{r}=\frac{1}{2}\rho _{0}\int
\left\vert \mathbf{u}\right\vert ^{2}\omega _{m}^{2}d^{3}\mathbf{r}=\rho
_{0}\left\vert u\right\vert ^{2}\omega _{m}^{2}\int \vec{\psi}^{\ast }\left(
\mathbf{r}\right) \cdot \vec{\psi}\left( \mathbf{r}\right) d^{3}\mathbf{r=}%
\frac{\hbar \omega _{m}}{2}\left\vert u\right\vert ^{2}$ and hence%
\begin{equation}
\rho _{0}\omega _{m}^{2}\int \vec{\psi}^{\ast }\left( \mathbf{r}\right)
\cdot \vec{\psi}\left( \mathbf{r}\right) d^{3}\mathbf{r}=\frac{\hbar \omega
_{m}}{2}.  \label{e26}
\end{equation}%
We can also introduce a dimensionless quantity $\left( 2\pi \frac{\hbar
\omega }{\epsilon }\right) \left\vert a\right\vert ^{2}=\left\vert \mathcal{E%
}\right\vert ^{2}$. In terms of the dimensionless quantities $a_{1}$, $a_{2}$
and $u$, the final equations are%
\begin{eqnarray}
&&\frac{\partial a_{2}}{\partial t}=-\kappa _{2}a_{2}-\text{i}\Delta
_{2}a_{2}-\text{i}\beta ua_{1}\mathrm{e}^{\mathrm{i}\delta t},  \label{e27}
\\
&&\frac{\partial a_{1}}{\partial t}=-\kappa _{1}a_{1}-\text{i}\Delta
_{1}a_{1}-\text{i}\beta ^{\ast }u^{\ast }a_{2}\mathrm{e}^{-\mathrm{i}\delta
t},  \label{e28} \\
&&\frac{\partial u}{\partial t}=-\Gamma u-\text{i}\beta ^{\ast }a_{1}^{\ast
}a_{2}\mathrm{e}^{-\mathrm{i}\delta t},  \label{e29}
\end{eqnarray}%
where the coupling constant $\beta $ is now given by%
\begin{equation}
\beta =\frac{\gamma _{e}}{2}\sqrt{\frac{\omega _{\mathrm{c}2}\omega _{%
\mathrm{c}1}}{\epsilon _{2}\epsilon _{1}}}\int \left[ \vec{\phi}_{2}^{\ast
}\left( \mathbf{r}\right) \cdot \vec{\phi}_{1}\left( \mathbf{r}\right) %
\right] \left( \nabla \cdot \vec{\psi}\left( \mathbf{r}\right) \right) d^{3}%
\mathbf{r}.  \label{e30a}
\end{equation}%
For a more general phonon mode described by the interaction density in Eq. (%
\ref{e14a}), one can follow a similar procedure as detailed above for the
longitudinal acoustic mode. Using the same symbols as in the equation (\ref%
{e17}), one can describe the general vibrational mode by its displacement
field%
\begin{equation}
\mathbf{Q}(\mathbf{r},t)=\vec{\psi}\left( \mathbf{r}\right) u\left( t\right)
e^{-\text{i}\omega _{m}t}+c.c,  \label{e30b}
\end{equation}%
where $u\left( t\right) $ is again the slowly varying dimensionless
amplitude of the phonon mode and $\vec{\psi}\left( \mathbf{r}\right) $ is
the corresponding mode function. Instead of introducing too many additional
symbols, note that we are using the same symbols for mode functions and
their amplitudes even for the general phonon mode. It requires a few words
of caution. It should always be remembered that now $\nabla \times \vec{\psi}
$ is not necessarily zero because it may not be representing a longitudinal
phonon mode.

In such a case, one can show that we obtain the same set of final working
equations (\ref{e27}) - (\ref{e29}), except that instead of the expression (%
\ref{e30a}), the anti-Stokes coupling constant is now defined by%
\begin{equation}
\beta ^{\left( Q\right) }=2\pi \sqrt{\frac{\omega _{\mathrm{c}2}\omega _{%
\mathrm{c}1}}{\epsilon _{2}\epsilon _{1}}}\sum_{ijk}R_{ijk}\int \phi
_{2i}^{\ast }\left( \mathbf{r}\right) \phi _{1j}\left( \mathbf{r}\right)
\psi _{k}\left( \mathbf{r}\right) d^{3}\mathbf{r}.  \label{e30c}
\end{equation}

It has to be noted that the resonator is being driven by a pump field of
frequency $\omega _{1}$. Thus Eq.(\ref{e27}) is to be modified by adding the
effect of the external field. We change $\kappa _{1}a_{1}$ term:%
\begin{equation}
\kappa _{1}a_{1}\rightarrow \kappa _{1}\left( a_{1}-\mathcal{\tilde{E}}%
_{1}\right) .  \label{e31}
\end{equation}%
Here $\mathcal{\tilde{E}}_{1}$ is dimensionless and is determined by the
resonator boundary conditions. The phonon mode is driven by thermal
fluctuations, thus Eq. (\ref{e29}) is to be modified by converting it into a
Langevin equation (see Eq. (\ref{ev5})). The Eq. (\ref{e28}) is also
converted to Langevin equation by adding terms corresponding to the input
vacuum field.

It is quite interesting to note that the basic equations (\ref{e27}) - (\ref%
{e29}) are quite generic in nature as these are applicable to a variety of
situations. The details of a specific system enter through the coupling
constant $\beta $ and the damping parameter $\Gamma ^{^{\prime }s}$ and $%
\kappa ^{^{\prime }s}$. The coupling constant Eq. (\ref{e30a}) depends on
the mode under consideration. We also note that in deriving equations (\ref%
{e27}) - (\ref{e29}) we have ignored the Stokes processes which are expected
to be unimportant as we assume that there is no mode resonant with the
Stokes frequency.

We also note that the coherent terms (i.e. terms without damping) in (\ref%
{e27}) - (\ref{e29}) can be obtained from the following effective Hamiltonian%
\begin{equation}
\mathcal{H}_{\text{eff}}=\hbar \Delta _{1}a_{1}^{\dagger }a_{1}+\hbar \Delta
_{2}a_{2}^{\dagger }a_{2}+\hbar \left( \beta ^{\ast }b^{\dagger
}a_{1}^{\dagger }a_{2}\mathrm{e}^{-\mathrm{i}\delta t}+H.c.\right) ,
\label{e36}
\end{equation}%
where we have used the second quantized notation $u\rightarrow b$, $u^{\ast
}\rightarrow b^{\dagger }$, $a_{i}^{\ast }\rightarrow a_{i}^{\dagger }$. The
operators $a_{i}^{^{\prime }s}$, $a_{i}^{^{\dagger \prime }s}$, $b$ and $%
b^{\dagger }$ satisfy Bosonic commutation relations. Under the assumption
that the pump $a_{1}$ remains undepleted and is at resonance with the cavity
mode ($\Delta _{1}\simeq 0$), $a_{1}\rightarrow \mathcal{E}_{1}$, Eq. (\ref%
{e36}) reduces to%
\begin{equation}
\mathcal{H}_{\text{eff}}=\hbar \Delta _{2}a_{2}^{\dagger }a_{2}+\hbar \left(
\beta ^{\ast }b^{\dagger }a_{2}\mathcal{E}_{1}^{\ast }\mathrm{e}^{-\mathrm{i}%
\delta t}+H.c.\right) .  \label{e37}
\end{equation}%
It should be borne in mind that the time dependence like $e^{-\text{i}\omega
_{2}t}$ and $e^{-\text{i}\omega _{m}t}$ were separated out from $a_{2}$ and $%
b$ (Eq. (\ref{e15}) and Eq. (\ref{e17})). If we work in a picture where such
time dependences are kept, then instead of Eq. (\ref{e37}), we will have%
\begin{equation}
\mathcal{H}_{\text{eff}}=\hbar \omega _{\mathrm{c}2}a_{2}^{\dagger
}a_{2}+\hbar \omega _{m}b^{\dagger }b+\hbar \left( \beta ^{\ast }b^{\dagger
}a_{2}\mathcal{E}_{1}^{\ast }\mathrm{e}^{\mathrm{i}\omega _{1}t}+H.c.\right)
.  \label{e38}
\end{equation}%
It is important to keep track of the fast time dependences. Having derived
Eq. (\ref{e38}) from first principles, we will use it's form to write the
corresponding Hamiltonian for the case of two phonon modes.

\section{PHONON DISPERSION RELATIONS , CAVITY MODES AND DAMPING PARAMETERS}

Let us first consider the general problem of cooling phonon modes in a bulk
material, which are characterized by the dispersion relation $\Omega \left(
\mathbf{q}\right) $, where $\mathbf{q}$ is the wave vector in a bulk
material. Here the mode functions are plane waves. The phonon field is
proportional to $\exp \left\{ \text{i}\mathbf{q}\cdot \mathbf{r}-\text{i}%
\Omega \left( \mathbf{q}\right) t\right\} $. We could have transverse
phonons as in the case of Raman active optical phonons in solids, or
longitudinal acoustic phonons as, say, in the case of Brillouin scattering.
As already discussed earlier, the physical mechanism for cooling is the
upconversion (anti-Stokes process) of light of frequency $\omega _{1}$ and
wave vector $\mathbf{k}_{1}$. The upconversion is subject to the energy
momentum conservation%
\begin{equation*}
\Omega \left( \mathbf{q}\right) +\omega _{1}\left( \mathbf{k}_{1}\right)
\rightarrow \omega _{2}\left( \mathbf{k}_{2}\right)
\begin{array}{cc}
, &
\end{array}%
\mathbf{q}+\mathbf{k}_{1}=\mathbf{k}_{2}
\end{equation*}%
In the upconversion process a thermal phonon of frequency $\Omega \left(
\mathbf{q}\right) $ is quickly removed. Further in order for the
upconversion to be effective, we have to make sure that the process of
downconversion (Stokes process) is avoided%
\begin{equation*}
\omega _{1}\left( \mathbf{k}_{1}\right) -\Omega \left( \mathbf{q}\right)
=\omega _{s}\left( \mathbf{k}_{s}\right)
\begin{array}{cc}
, &
\end{array}%
\mathbf{k}_{1}-\mathbf{q}=\mathbf{k}_{s}
\end{equation*}%
This is where the use of resonators and cavities becomes important. The form
of the dispersion relation or the frequency spectrum of the phonon modes is
also important. Several interesting forms of dispersion relations in simple
geometries can be noted:

\begin{enumerate}
\item Bulk longitudinal sound waves - Brillouin mode: $\Omega \left(
q\right) =v_{s}q$, where $v_{s}$ is the velocity of sound waves;

\item Bulk Raman active optical modes: \ $\Omega ^{2}\left( q\right) =\Omega
_{0}^{2}-\alpha q^{2}$;

\item Modes in confined geometries: it is well known that if the sample is
confined to a length $d$ in any given direction, phonon modes propagating in
that direction are discrete with spacing proportional to $\pi /d$. If the
sample is confined in two of its dimensions, as in the case of phonons in an
optical fiber, for a given discrete phonon mode in the transverse direction,
the phonon frequency $\Omega \left( q\right) $ in the propagating direction
is given by $\Omega ^{2}\left( q\right) \approx \Omega _{0}^{2}+\alpha q^{2}$%
, where $\Omega _{0}$ for different discrete modes in the transverse
direction is determined by the corresponding zero of the Bessel functions
and thus it is inversely proportional to the radius of the fiber. In cases
when $\Omega _{0}\neq 0$, the $q$ dependence of the frequency is not very
significant and then the condition for momentum conservation in the
direction of propagation is of a minor consequence. In a spherical
resonator, which is confined in all the three directions, modes (both radial
and azimuthal) are discrete.
\end{enumerate}

\begin{figure}[t]
\includegraphics[width=0.49\textwidth]{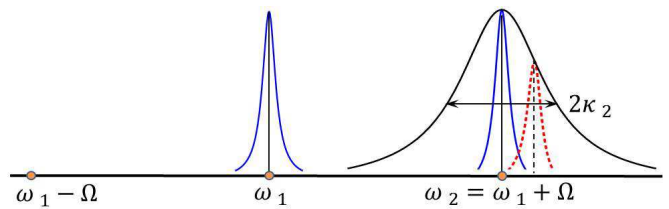}
\caption{Schematic representation of the anti-Stokes generation and the
width of varies lines.}
\label{Fig1}
\end{figure}

One can arrange the resonator structure such that the upconversion process
is resonant with the cavity mode whereas the downconversion process is
non-resonant. Further one can choose the resonant cavity mode at $\omega
_{2} $ to have large damping $\kappa _{2}$ so that the generated anti-Stokes
field is quickly removed from the cavity. We assume that the width $\Gamma
_{m}$ of the phonon mode is much smaller than the width of the cavity mode
at the generated anti-Stokes frequency. We also assume that the pump laser
at $\omega _{1}$ is monochromatic and that the line width of the
corresponding resonator mode at $\omega _{1}$ is much smaller than $\kappa
_{2}$. The situation is shown in the Fig.\ref{Fig1} The anti-Stokes phonon
could be exactly at the line center of the resonator mode at $\omega _{2}$.
It depends on the energy conservation condition.
The Figure \ref{Fig1} also implies that more than one phonon mode (see for
example, the dotted curve) can be cooled simultaneously as long as these lie
well within the width $2\kappa _{2}$ of the generated anti-Stokes field; in
other words if $\Omega _{1}$ and $\Omega _{2}$ are the center frequencies of
two nearby sharp phonon modes, one should have $|\Omega _{1}-\Omega
_{2}|<2\kappa _{2}$. We consider this in detail in the next section.

\section{Theoretical Description of Bimodal Cooling}

Let us then first consider the case of two phonons modes in a resonator, one
with frequency $\Omega _{1}$ and the other with frequency $\Omega _{2}$. Let
the pump laser be at the frequency $\omega _{1}$. Then as required by the
Fig.\ref{Fig1}, we assume that $\omega _{1}+$ $\Omega _{i}$ $\left(
i=1,2\right) $ lie within the cavity line width $\kappa _{2}$ where the
cavity is centered at the frequency $\omega _{\mathrm{c}}$. Under the
assumption that the pump field remains un-depleted, the Hamiltonian for a
system of two phonon modes $b_{1}$, $b_{2}$ and the cavity mode $a_{2}$ can
be written as%
\begin{eqnarray}
H &=&\hbar \omega _{\mathrm{c}}a_{2}^{\dagger }a_{2}+  \label{ev1} \\
&&+\sum_{i=1,2}\left\{ \hbar \Omega _{i}b_{i}^{\dagger }b_{i}+\left( \beta
_{i}^{\ast }b_{i}^{\dagger }a_{2}\mathcal{E}_{1}^{\ast }\mathrm{e}^{-\mathrm{%
i}\omega _{1}t}+H.c.\right) \right\} ,  \notag
\end{eqnarray}%
where we have replaced cavity mode $a_{1}$ by $\mathcal{E}_{1}$. The mode
operators $a_{2}$, $b_{1}$ and $b_{2}$ satisfy Bosonic commutation
relations. We assume that the cavity field is built up from the interaction
of the pump field with thermal phonons. Let $\Omega _{0}$ be $\frac{\Omega
_{1}+\Omega _{2}}{2}$ and $\Omega =\frac{\Omega _{1}-\Omega _{2}}{2}$, then
we work in a picture obtained by making unitary transformation with
\begin{equation}
\bar{H}=\hbar \left( \omega _{1}+\Omega _{0}\right) a_{2}^{\dagger
}a_{2}+\hbar \Omega _{0}\left( b_{1}^{\dagger }b_{1}+b_{2}^{\dagger
}b_{2}\right)  \label{ev2}
\end{equation}%
to obtain an effective Hamiltonian%
\begin{eqnarray}
H_{\text{eff}} &=&\hbar \Omega \left( b_{1}^{\dagger }b_{1}-b_{2}^{\dagger
}b_{2}\right) +\hbar \delta a_{2}^{\dagger }a_{2}  \notag \\
&&+\sum_{i=1,2}\left( \beta _{i}^{\ast }b_{i}^{\dagger }a_{2}\mathcal{E}%
_{1}^{\ast }\mathrm{e}^{\mathrm{i}\delta t}+H.c.\right)  \notag \\
&=&\hbar \Omega \left( b_{1}^{\dagger }b_{1}-b_{2}^{\dagger }b_{2}\right)
+\hbar \delta a_{2}^{\dagger }a_{2}  \notag \\
&&+\sum_{i=1,2}\left( G_{i}^{\ast }b_{i}^{\dagger }a_{2}\mathrm{e}^{\mathrm{i%
}\delta t}+H.c.\right) ,  \label{ev3} \\
\delta &=&\omega _{\mathrm{c}}-\left( \omega _{1}+\Omega _{0}\right)
\begin{array}{cc}
, &
\end{array}%
G_{i}=\beta _{i}\mathcal{E}_{1}.  \label{ev4}
\end{eqnarray}%
This is our working Hamiltonian which we use to obtain Heisenberg equations
for $a_{2}$, $b_{i}$. We now need to introduce the dissipative terms which
account for the leakage of phonons for the resonator and dissipation of the
phonon modes and their then temperature. It is best then to write the
quantum Langevin equations for the Heisenberg operators $a_{2}$ and $b_{i}$:%
\begin{eqnarray}
&&\dot{a}_{2}=-\text{i}\delta a_{2}-\kappa _{2}a_{2}-\text{i}\left(
G_{1}b_{1}+G_{2}b_{2}\right) +f_{a_{2}}\left( t\right) ,  \notag \\
&&\dot{b}_{1}=-\text{i}\Omega b_{1}-\Gamma _{1}b_{1}-\text{i}G_{1}^{\ast
}a_{2}+f_{b_{1}}\left( t\right) ,  \notag \\
&&\dot{b}_{2}=+\text{i}\Omega b_{2}-\Gamma _{2}b_{2}-\text{i}G_{2}^{\ast
}a_{2}+f_{b_{2}}\left( t\right) .  \label{ev5}
\end{eqnarray}%
Here $\Gamma _{i}$ denotes the half width of the phonon mode $i$, $2\kappa
_{2}$ denotes the line width of the cavity mode.

Let us note here that the above set of equations are very generic, which can
describe the coupling of any two-mode system to a single optical field. In
this sense, the resulting physics here in our case should be similar to the
studies for other systems \cite{ar6,ar7,ar8,ar9} involving these equations,
including systems of oscillating membranes and trapped atoms.

In what follows, we would work in the limit $\Gamma _{i}\ll \kappa _{2}$; $%
\Omega _{i}\gg \kappa _{2}$. The force term \ $f_{j}\left( t\right) $ are
the quantum Langevin forces. These are Gaussian forces with zero mean and
with quantum correlations given by%
\begin{eqnarray}
\left\langle f_{a_{2}}^{\dagger }\left( t\right) f_{a_{2}}\left( t^{\prime
}\right) \right\rangle  &=&0%
\begin{array}{cc}
, &
\end{array}%
\left\langle f_{a_{2}}\left( t\right) f_{a_{2}}^{\dagger }\left( t^{\prime
}\right) \right\rangle =2\kappa _{2}\delta \left( t-t^{\prime }\right) ,
\notag \\
\left\langle f_{b_{i}}^{\dagger }\left( t\right) f_{b_{i}}\left( t^{\prime
}\right) \right\rangle  &=&\frac{2\Gamma _{i}}{\pi }\int_{-\infty }^{\infty }%
\bar{n}_{i}\left( \omega \right) \mathrm{e}^{-\mathrm{i}\omega \left(
t-t^{\prime }\right) }d\omega ,  \notag \\
\left\langle f_{b_{i}}\left( t\right) f_{b_{i}}^{\dagger }\left( t^{\prime
}\right) \right\rangle  &=&\frac{2\Gamma _{i}}{\pi }\int_{-\infty }^{\infty
}\left( \bar{n}_{i}\left( \omega \right) +1\right) \mathrm{e}^{-\mathrm{i}%
\omega \left( t-t^{\prime }\right) }d\omega ,  \notag \\
\left\langle f_{b_{1}}\left( t\right) f_{b_{2}}^{\dagger }\left( t^{\prime
}\right) \right\rangle  &=&0.  \label{ev6}
\end{eqnarray}%
A detailed derivation of the correlation functions in Eq. (\ref{ev6}) can be
found in books on quantum optics \cite{bk3}.

We first analyze the well known case \cite{ar2}\ of the cooling of a single
mode in the limit of large $\kappa _{2}$. We can then make the adiabatic
approximation and write%
\begin{equation}
a_{2}\approx \left( \kappa _{2}+\text{i}\delta \right) ^{-1}\left( -\text{i}%
G_{1}\right) b_{1},  \label{ev7}
\end{equation}%
and then
\begin{equation*}
\frac{\partial b_{1}}{\partial t}+\text{i}\Omega b_{1}+\left( \Gamma _{1}+%
\frac{\left\vert G_{1}\right\vert ^{2}}{\kappa _{2}+\text{i}\delta }\right)
b_{1}=f_{b_{1}},
\end{equation*}%
i.e.%
\begin{eqnarray}
&&\frac{\partial b_{1}}{\partial t}+\text{i}\Omega _{1\text{eff}%
}b_{1}+\Gamma _{1\text{eff}}b_{1}=f_{b_{1}},  \label{ev8} \\
&&\Gamma _{1\text{eff}}=\left( \Gamma _{1}+\frac{\left\vert G_{1}\right\vert
^{2}\kappa _{2}}{\kappa _{2}^{2}+\delta ^{2}}\right) ,  \notag \\
&&\Omega _{1\text{eff}}=\Omega _{1}-\frac{\delta \left\vert G_{1}\right\vert
^{2}}{\kappa _{2}^{2}+\delta ^{2}}.  \label{ev9}
\end{eqnarray}%
Thus the life time of the phonon mode goes down as $\Gamma _{1\text{eff}}$
increases. The increase of $\Gamma _{1\text{eff}}$ is subject to the
assumption $\Gamma _{1\text{eff}}<\kappa _{2}$. The increase of $\Gamma $
leads to the cooling of the phonon mode, as the strength of the fluctuation $%
f_{b_{1}}$ is still determined by Equations in (\ref{ev6}), i.e. by the
damping parameter $\Gamma _{1}$. Note further that the fluctuation $f_{a_{2}}
$ associated with the field mode does not contribute to the normally ordered
moments as $\left\langle f_{a_{2}}^{\dagger }\left( t\right) f_{a_{2}}\left(
t^{\prime }\right) \right\rangle =0$, which is the case as long as $\hbar
\omega /k_{B}T\gg 1$.

\section{Mode-Mode Coupling and the Collective Behavior of the phonon Modes}

We will now show that the adiabatic elimination of the cavity field results
in mode-mode coupling. The resulting Langevin equations for phonon modes are
given by%
\begin{equation*}
\frac{\partial b_{1}}{\partial t}+\left( \Gamma _{1}+\text{i}\Omega \right)
b_{1}=-\frac{\left\vert G_{1}\right\vert ^{2}b_{1}+G_{1}^{\ast }G_{2}b_{2}}{%
\kappa _{2}+\text{i}\delta }+f_{b_{1}}\left( t\right) ,
\end{equation*}%
\begin{equation}
\frac{\partial b_{2}}{\partial t}+\left( \Gamma _{2}-\text{i}\Omega \right)
b_{2}=-\frac{\left\vert G_{2}\right\vert ^{2}b_{2}+G_{2}^{\ast }G_{1}b_{1}}{%
\kappa _{2}+\text{i}\delta }+f_{b_{2}}\left( t\right) .  \label{ev10}
\end{equation}%
The mode-mode coupling can be thought of as the process%
\begin{eqnarray*}
&&\text{phonon }1+\text{pump field}\rightarrow \text{anti-Stokes field} \\
&\rightarrow &\text{phonon }2+\text{pump field}
\end{eqnarray*}%
It should be borne in mind that in the picture we work, all frequencies are
matched due to Eq.(\ref{ev2}). The actual generation of $b_{2}$ depends on
the frequency mismatch factors $\Omega $ and $\delta $ as the solution of
equations in (\ref{ev10}) will obviously show.

Let us now consider the possibility of the collective behavior of the phonon
modes. Let there be two identical phonon modes. This would be the case for
two identical nano-mirrors placed at different nodes of an optical cavity
(Similar to the situation of ref. \cite{ar6}); We can then set $\Omega =0$.
We drop $\Gamma _{i}$ assuming that the coupling field strength is such that
$\Gamma _{\text{eff}}\sim \frac{\left\vert G_{1}\right\vert ^{2}\kappa _{2}}{%
\kappa _{2}^{2}+\delta ^{2}}$. Then equations in (\ref{ev10}) reduce to%
\begin{eqnarray}
\frac{\partial b_{1}}{\partial t} &\approx &-\frac{\left\vert
G_{1}\right\vert ^{2}b_{1}+G_{1}^{\ast }G_{2}b_{2}}{\kappa _{2}+\text{i}%
\delta }+f_{b_{1}}\left( t\right) ,  \notag \\
\frac{\partial b_{2}}{\partial t} &\approx &-\frac{\left\vert
G_{2}\right\vert ^{2}b_{2}+G_{2}^{\ast }G_{1}b_{1}}{\kappa _{2}+\text{i}%
\delta }+f_{b_{2}}\left( t\right) .  \label{ev12}
\end{eqnarray}%
The two noise sources would also be identical. Further under the simplifying
assumption $G_{1}=G_{2}=$real%
\begin{eqnarray}
\frac{\partial b_{+}}{\partial t} &=&-2\frac{G_{1}^{2}b_{+}}{\kappa _{2}+%
\text{i}\delta }+f_{+}\left( t\right) ,  \notag \\
\frac{\partial b_{-}}{\partial t} &=&0+f_{-}\left( t\right) =0,  \label{ev13}
\\
b_{\pm } &=&\frac{b_{1}\pm b_{2}}{\sqrt{2}}%
\begin{array}{cc}
, &
\end{array}%
f_{\pm }=\frac{f_{b_{1}}\pm f_{b_{2}}}{\sqrt{2}}.  \notag
\end{eqnarray}%
Under the conditions $\Gamma _{1}=\Gamma _{2}$, $\Omega =0$ used in deriving
Eq. (\ref{ev13}), $f_{-}$ does not lead to nonvanishing contribution. Thus
the collective mode $b_{-}$ is sub-radiant whereas the mode $b_{+}$ is
super-radiant. The collective mode $b_{+}$ has a decay rate which is twice
than what it would be in the absence of any collective effects. Under these
very special conditions there is enhanced cooling of only the collective
mode $b_{+}$ where $b_{-}$ mode is unaffected. The collective effects are
known in other contexts, for example in the cooling of two trapped ions \cite%
{ar7} and in electro mechanical oscillators \cite{ar6}. The collective modes
in an ensemble of mechanical resonators trapped atoms in a Fabry-Perot
resonator have been studied \cite{ar8,ar9}.

\section{Spectrum of the Generated Anti-Stokes Field and Cooling of the
Phonon Modes}

\begin{figure}[t]
\includegraphics[width=0.49\textwidth]{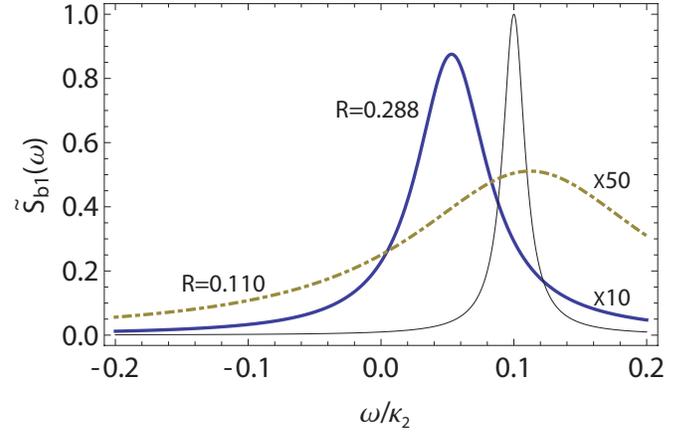}
\caption{The spectrum $\tilde{S}_{b_{1}}\left( \protect\omega \right) =\frac{%
\Gamma _{1}S_{b_{1}}\left( \protect\omega \right) }{2n_{1}}$ as a function
of $\protect\omega /\protect\kappa _{2}$ for $\frac{\Gamma _{1}}{\protect%
\kappa _{2}}=\frac{\Gamma _{1}}{\protect\kappa _{2}}=0.01$; $\frac{\Omega }{%
\protect\kappa _{2}}=0.1$ and $\frac{G_{1}}{\protect\kappa _{2}}=0.3$, $%
G_{2}=0$ (green curve), $\frac{G_{1}}{\protect\kappa _{2}}=0.3$, $\frac{G_{2}%
}{\protect\kappa _{2}}=0.5$ (blue curve). The black curve represents
spectrum when both $G_{1}=G_{2}=0$. The parameter $R$ (Eq. \protect\ref{ev19}%
) gives the extent of cooling.}
\label{Fig2}
\end{figure}
The spectrum of phonon fluctuations can be obtained by solving the quantum
Langevin equations (\ref{ev5}). Defining Fourier transformations via%
\begin{equation}
b\left( t\right) =\frac{1}{2\pi }\int_{-\infty }^{\infty }b\left( \omega
\right) \mathrm{e}^{-\mathrm{i}\omega t}d\omega ,  \label{ev14}
\end{equation}%
so that%
\begin{equation}
\left\langle b^{\dagger }\left( t\right) b\left( t\right) \right\rangle =%
\frac{1}{2\pi }\int_{-\infty }^{\infty }S_{b}\left( \omega \right) d\omega ,
\label{ev15}
\end{equation}%
\begin{equation}
2\pi S_{b}\left( \omega \right) \delta \left( \omega ^{\prime }+\omega
\right) =\left\langle b^{\dagger }\left( -\omega ^{\prime }\right) b\left(
\omega \right) \right\rangle .  \label{ev16}
\end{equation}

The quantity $S_{b}\left( \omega \right) $ gives the spectrum of phonon
fluctuations. A reduction in the value of $\left\langle b^{\dagger
}b\right\rangle $ when $G^{\prime s}\neq 0$, would represent cooling. Using
the properties of the Langevin forces we find the final results for the
spectra of the two phonon modes
\begin{subequations}
\label{ev17}
\begin{align}
S_{b_{1}}\left( \omega \right) =& \frac{2\Gamma _{1}n_{1}\left( \omega
\right) \left\vert \left( \mathrm{i}\delta -\mathrm{i}\omega +\kappa
_{2}\right) +\frac{\left\vert G_{2}\right\vert ^{2}}{-\mathrm{i}\Omega -%
\mathrm{i}\omega +\Gamma _{2}}\right\vert ^{2}}{\left\vert d\left( \omega
\right) \right\vert ^{2}\left[ \left( \Omega -\omega \right) ^{2}+\Gamma
_{1}^{2}\right] }  \notag \\
& +\frac{2\Gamma _{2}n_{2}\left( \omega \right) \left\vert \frac{G_{1}G_{2}}{%
-\mathrm{i}\Omega -\mathrm{i}\omega +\Gamma _{2}}\right\vert ^{2}}{%
\left\vert d\left( \omega \right) \right\vert ^{2}\left[ \left( \Omega
-\omega \right) ^{2}+\Gamma _{1}^{2}\right] }  \notag \\
& \underrightarrow{G_{1},G_{2}=0}\text{ }\frac{2\Gamma _{1}n_{1}\left(
\omega \right) }{\left[ \left( \Omega -\omega \right) ^{2}+\Gamma _{1}^{2}%
\right] },  \label{ev17a}
\end{align}%
\begin{align}
S_{b_{2}}\left( \omega \right) =& \frac{2\Gamma _{2}n_{2}\left( \omega
\right) \left\vert \left( \mathrm{i}\delta -\mathrm{i}\omega +\kappa
_{2}\right) +\frac{\left\vert G_{1}\right\vert ^{2}}{-\mathrm{i}\Omega -%
\mathrm{i}\omega +\Gamma _{1}}\right\vert ^{2}}{\left\vert d\left( \omega
\right) \right\vert ^{2}\left[ \left( \Omega +\omega \right) ^{2}+\Gamma
_{2}^{2}\right] }  \notag \\
& +\frac{2\Gamma _{1}n_{1}\left( \omega \right) \left\vert \frac{G_{1}G_{2}}{%
\mathrm{i}\Omega -\mathrm{i}\omega +\Gamma _{1}}\right\vert ^{2}}{\left\vert
d\left( \omega \right) \right\vert ^{2}\left[ \left( \Omega +\omega \right)
^{2}+\Gamma _{2}^{2}\right] }  \notag \\
& \underrightarrow{G_{1},G_{2}=0}\text{ }\frac{2\Gamma _{2}n_{2}\left(
\omega \right) }{\left[ \left( \Omega +\omega \right) ^{2}+\Gamma _{1}^{2}%
\right] },  \label{ev17b}
\end{align}%
\begin{figure}[bt]
\includegraphics[width=0.49\textwidth]{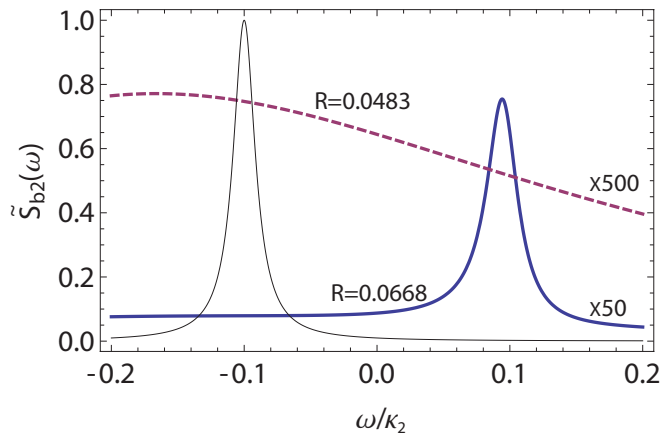}
\caption{Same as in Fig.\protect\ref{Fig2}, except for the spectrum $\tilde{S%
}_{b_{2}}\left( \protect\omega \right) =\frac{\Gamma _{2}S_{b_{2}}\left(
\protect\omega \right) }{2n_{2}}$ of the $b$ mode. The pink graph is for $%
G_{1}=0$, $\frac{G_{2}}{\protect\kappa _{2}}=0.5$.}
\label{Fig3}
\end{figure}
where
\end{subequations}
\begin{equation}
d\left( \omega \right) =\left( \mathrm{i}\delta -\mathrm{i}\omega +\kappa
_{2}\right) +\frac{\left\vert G_{1}\right\vert ^{2}}{\mathrm{i}\Omega -%
\mathrm{i}\omega +\Gamma _{1}}+\frac{\left\vert G_{2}\right\vert ^{2}}{-%
\mathrm{i}\Omega -\mathrm{i}\omega +\Gamma _{2}}.  \label{ev18}
\end{equation}

The spectrum of fluctuations, given by equations in (\ref{ev17}), should be
compared with the unperturbed spectrum ($G_{1}=G_{2}=0$). For typical
parameters we show the results in Figs. \ref{Fig2} and \ref{Fig3}. The
parameters $\kappa _{2}$, $\Gamma ^{^{\prime }s}$ and $G^{^{\prime }s}$ that
we choose, are similar to the case of a single phonon mode \cite{ar5}.
Although the coupling of the three modes makes $d\left( \omega \right) $
cubic function of $\omega $, the spectrum in the vicinity of $\omega =\Omega
$ ($\omega =-\Omega $) exhibits a single resonance for $\delta =0$, $\Gamma
\ll \kappa $. The Figs. \ref{Fig2} and \ref{Fig3} show the cooling of both
modes. The mode-mode coupling leads to heating, however overall both modes
get cooled. The ratio $R$ gives the ratio of $\left\langle b^{\dagger
}\left( t\right) b\left( t\right) \right\rangle $ in presence of the
coupling to it's value in the absence of coupling i.e. $R$ for example in
Fig. \ref{Fig2} is%
\begin{equation}
R=\frac{\left\langle b_{1}^{\dagger }\left( t\right) b_{1}\left( t\right)
\right\rangle }{\bar{n}_{1}}.  \label{ev19}
\end{equation}%
It should be emphasized that the cooling parameter $R$ introduced here
should not be confused with the effective bulk Raman coupling constant $%
R^{\left( Q\right) }$ introduced in Eq. (\ref{e14b}). For Fig.\ref{Fig2} for
the values of $G_{1}$ and $G_{2}$ used, $R=0.110$ if $G_{2}=0$; $R=0.288$ if
$G_{2}\neq 0$. The parameters $G_{1}$ and $G_{2}$ [Eq. (\ref{ev4})] depend
on the phase matching parameter $\beta $ and the power of the Stokes field.
Clearly if we increase the power of the Stokes laser i.e. if we increase $%
G_{,}$ then $R$ goes down. We note that the parameter $R$ also gives the
ratio of the temperature of the cooled phonon mode to its initial
temperature. This is because the factor $\left( e^{\hbar \omega /k_{\text{B}%
}T}-1\right) ^{-1}\approx \frac{k_{\text{B}}T}{\hbar \omega }$ if $\frac{k_{%
\text{B}}T}{\hbar \omega }\ll 1$. For phonon frequency, say in the range, $%
100$MHz \cite{ar5}, the parameter $\frac{k_{\text{B}}T}{\hbar \omega }\ll 1$
even at a temperature like $1$K.

\begin{figure}[t]
\vspace{.5cm} \includegraphics[width=0.49\textwidth]{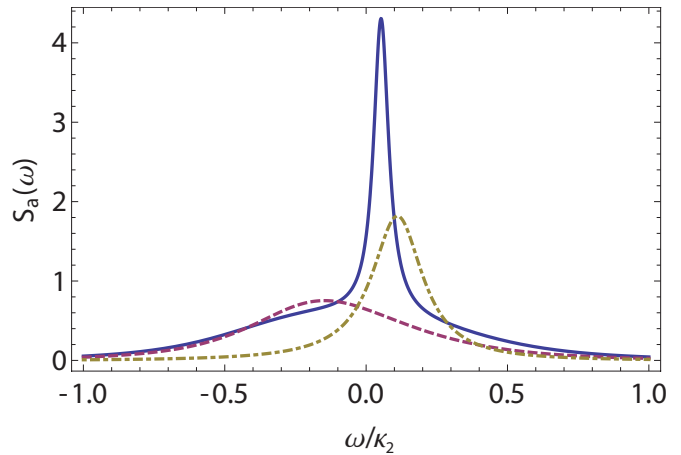}
\caption{Spectrum of the generated anti-Stokes for $\frac{G_{1}}{\protect%
\kappa _{2}}=0.3$, $G_{2}=0$ (green); $G_{1}=0$, $\frac{G_{2}}{\protect%
\kappa _{2}}=0.5$ (pink); $\frac{G_{1}}{\protect\kappa _{2}}=0.3$, $\frac{%
G_{2}}{\protect\kappa _{2}}=0.5$ (blue).}
\label{Fig4}
\end{figure}
The spectrum of the generated anti-Stokes field is given by
\begin{equation}
S_{a}\left( \omega \right) =\frac{2\Gamma _{1}n_{1}\left( \omega \right)
\left\vert \frac{G_{1}}{\mathrm{i}\Omega -\mathrm{i}\omega +\Gamma _{1}}%
\right\vert ^{2}+2\Gamma _{2}n_{2}\left( \omega \right) \left\vert \frac{%
G_{2}}{-\mathrm{i}\Omega -\mathrm{i}\omega +\Gamma _{2}}\right\vert ^{2}}{%
\left\vert d\left( \omega \right) \right\vert ^{2}},  \label{ev20}
\end{equation}%
which is shown in Fig.\ref{Fig4}. The figure also shows the effect of mode
mixing. The mode mixing arises from the modification of the denominator in
Eq. (\ref{ev20}). The numerator in Eq. (\ref{ev20}) represents the
conversion of phonons into anti-Stokes in the lowest order of the coupling $%
G_{1}$ and $G_{2}$. Note that the spectrum of the generated antiStokes
radiation has signatures of the phonon spectrum. The spectrum of the
generated radiation can be studied by examining the output field at
frequency $\omega _{2}$. Note that the anti-Stokes photons leak out of the
cavity and the phonon modes are constantly interacting with the thermal
bath, the conservation of photon and phonon numbers in the cavity does not
hold.

If we use the actual numerical parameters of the experiment on the single-
mode cooling of the acoustic phonon modes done by Bahl \textit{et al.} \cite%
{ar5}, we can also examine the possibility of bimodal phonon cooling in such
a system. In their experiment, $\kappa _{2}$ was of the order of $5$MHz, and
they had observed cooling of the surface acoustic phonon mode of frequency $%
95$MHz, with their intrinsic line width of the order of only $8$kHz at the
ambient temperature of $294$K. Thus, if one can find another similar phonon
mode in the system with its frequency within a few MHz away from the $95$MHz
mode, say in the range of $90$ to $100$MHz, which also has an appreciable
coupling $\beta $, defined by Eq. (\ref{e30a}), we should be able to observe
simultaneous cooling of two such modes in the same experimental system.
Otherwise, one has to search for another appropriate phonon system to
observe this effect. Note that in the experimental system discussed above,
if there are more than two phonon modes near the $95$MHz mode within $%
2\kappa _{2}$ range, each one with appreciable coupling, all those modes are
expected to be cooled simultaneously.

In conclusion, we have presented a general theory of cooling of phonon modes
via their interactions with suitable optical fields. The theory is
formulated in terms of the spatial mode functions for the relevant
interacting modes, in any given geometry, allowing it to be applicable to a
variety of systems with diverse phonon dispersion relations and frequency
spectra. In particular, we show the cooling of two phonon modes at the same
time if the line width of each of the phonon modes is very small compared to
the decay rate $2\kappa _{2}$ of the resonant anti-Stokes cavity mode and
their frequency difference is smaller than the line width $2\kappa _{2}$. In
any experimental set-up similar to that used by Bahl \textit{et al.} \cite%
{ar5}, if one can find two nearby sharp phonon modes satisfying the above
conditions , one should be able to observe the bimodal cooling described in
this paper. We also find in this paper the possibility of collectively
enhanced cooling of phonon modes under suitable conditions. However, further
studies are needed to identify such a system with required phonon mode
characteristics. A generalization of the present work to four wave
parametric interactions (for example Hamiltonians which are quadratic in $%
b_{i}^{^{\prime }s}$) would also be interesting.

One of us (SSJ) acknowledges the hospitality of the Oklahoma State
University, while this work was done.

\end{document}